\documentclass[letter]{aa} 
\usepackage{graphics,graphicx}
\usepackage{soul}
\usepackage{txfonts}
\usepackage{natbib}
\begin{document}
\newcommand{\phn}     {\phantom{0}}
\newcommand{\phnn}    {\phantom{0}\phantom{0}}
\newcommand{\phnnn}   {\phantom{0}\phantom{0}\phantom{0}}
\newcommand{\phs}     {\phantom{$>$}}
\newcommand{\et}      {et al.}
\newcommand{\ie}      {i.\,e.,}
\newcommand{\eg}      {e.\,g.,}
\newcommand{\mum}     {$\mu$m}
\newcommand{\kms}     {km~s$^{-1}$}
\newcommand{\cmt}     {cm$^{-3}$}
\newcommand{\jpb}     {$\rm Jy~beam^{-1}$}	
\newcommand{\nh}      {NH$_3$}
\newcommand{\nth}     {N$_2$H$^+$}
\newcommand{\Td}      {$T_\mathrm{d}$}
\newcommand{\Tex}     {T_\mathrm{ex}}
\newcommand{\Trot}    {T_\mathrm{rot}}
\newcommand{\cmg}     {cm$^{2}$~g$^{-1}$}
\newcommand{\chtoh}   {CH$_3$OH}
\newcommand{\water}   {H$_2$O}
\newcommand{\juc}     {\mbox{$J$=1$\rightarrow$0}}
\newcommand{\J}[2]    {\mbox{#1--#2}}
\newcommand{\uchii}   {UC~H{\small II}}
\newcommand{\hchii}   {HC~H{\small II}}
\newcommand{\raun}    {$^\mathrm{h~m~s}$}
\newcommand{\deun}    {$\mathrm{\arcdeg~\arcmin~\arcsec}$} 
\newcommand{\lo}      {$L_{\sun}$}
\newcommand{\mo}      {$M_{\sun}$}    
\newcommand{\Halpha}  {H110$\alpha$} 
\newcommand{\Calpha}  {C110$\alpha$} 
\newcommand{\hii}     {H{\small II}}
\newcommand{\hi}      {H{\small I}}
\newcommand{\oi}      {O{\small I}}
\newcommand{\oii}     {O{\small II}}
\newcommand{\ci}      {C{\small I}}
\newcommand{\ciii}    {C{\small II}}
\newcommand{\cii}     {C$^+$}
\newcommand{\MonR}    {Mon~R2}
%
\title{The first CO$^+$ image}
   \subtitle{I. Probing the \hi/H$_2$ layer around the ultracompact \hii\ region \MonR} 
\author{S.~P.~Trevi\~no-Morales\inst{1,2},
           A.~Fuente\inst{2}, 
           \'A.~S\'anchez-Monge\inst{3}, 
           P.~Pilleri\inst{4,5}, 
           J.~R.~Goicoechea\inst{1}, 
           V.~Ossenkopf-Okada\inst{3}, 
           E.~Roueff\inst{5}, 
           J.~R.~Rizzo\inst{6}, 
           M.~Gerin\inst{5}, 
           O.~Bern\'e\inst{4,7},
           J.~Cernicharo\inst{1}, 
           M.~G\'onzalez-Garc\'ia\inst{8}, 
           C.~Kramer\inst{9},
           S.~Garc\'ia-Burillo\inst{2} 
           J.~Pety\inst{10}}
           \authorrunning{S.~P.~Trevi\~no-Morales, 
           A.~Fuente, 
           \'A.~Sanchez-Monge, 
           et al.
           }

    \institute{Instituto de Ciencia de Materiales de Madrid, Sor Juana In\'es de la Cruz 3, E-28049 Cantoblanco, Madrid, (Spain)
    \and 	     
    Observatorio Astron\'omico Nacional, Apdo. 112, E-28803 Alcal\'a de Henares Madrid, (Spain) 
    \and
    I. Physikalisches Institut, Universit\"at zu K\"oln, Z\"ulpicher Str. 77, 50937 K\"oln, (Germany) 
    \and         
    CNRS; IRAP; 9 Av. colonel Roche, BP 44346, F-31028 Toulouse cedex 4, (France)
    \and         
    LERMA, Observatoire de Paris, PSL Research University, CNRS, UMR8112, Place Janssen, 92190 Meudon Cedex, (France)
    \and         
    Centro de Astrobiolog\'ia, E-28850 Torrej\'on de Ardoz, (Spain)
    \and     
    Universit\'e de Toulouse, UPS-OMP, IRAP, 31000 Toulouse, (France)
    \and         
    Instituto de Astrof\'isica de Andaluc\'ia, CSIC, E-18008, Granada, (Spain)
    \and
    Instituto de Radioastronom\'ia Milim\'etrica, Ave. Divina Pastora, 7, Local 20 18012, Granada (Spain)          
    \and         
    Institut de Radioastronomie Millim\'etrique, 300 Rue de la Piscine, F-38406 Saint Martin d'H\`eres, (France)
    }
         
    \date{Received ????; accepted ????}

    \abstract 
{The CO$^+$ reactive ion is thought to be a tracer of the boundary between a \hii\ region and the hot molecular gas. In this study, we present the spatial distribution of the CO$^+$ rotational emission toward the \MonR\ star-forming region. The CO$^+$ emission presents a clumpy ring-like morphology, arising from a narrow dense layer around the \hii\ region. We compare the CO$^+$ distribution with other species present in photon-dominated regions (PDR), such as [\ciii] 158~$\mu$m, H$_2$ S(3) rotational line at 9.3~$\mu$m, polycyclic aromatic hydrocarbons (PAHs) and HCO$^+$. We find that the CO$^+$ emission is spatially coincident with the PAHs and [\ciii] emission. This confirms that the CO$^+$ emission arises from a narrow dense layer of the \hi/H$_2$ interface. We have determined the CO$^+$ fractional abundance, relative to \cii\, toward three positions. The abundances range from 0.1 to 1.9 $\times 10^{-10}$ and are in good agreement with previous chemical model, which predicts that the production of CO$^+$ in PDRs only occurs in dense regions with high UV fields. The CO$^+$ linewidth is larger than those found in molecular gas tracers, and their central velocity are blue-shifted with respect to the molecular gas velocity. We interpret this as a hint that the CO$^+$ is probing photo-evaporating clump surfaces.}

 
\maketitle

\section{Introduction\label{s:intro}}

Reactive ions are destroyed in almost every collision with H and H$_2$ and recombine rapidly with e$^-$. These compounds present enhanced abundances toward the hot layers of the photon-dominated regions (PDRs), where the far ultraviolet (FUV) field is only partially attenuated and maintains high abundances of the parent species \cii\ and S$^+$ \citep{sternberg-Dalgarno1995}. In particular, \citet{sternberg-Dalgarno1995} predict a high CO$^+$ abundance at the \hi/H$_2$ interface ($A_\mathrm{V}\approx1$~mag) of dense PDRs, where it is mainly produced by the \cii\ + OH $\rightarrow$ CO$^+$ + H reaction. So far, the CO$^+$ ion has been detected in several PDRs; such as the M17SW, Orion Bar, NGC7027, NGC7023 (\citealt{latter1993}; \citealt{stoerzer1995}; \citealt{fuente1997}; \citealt{fuente2003}), G29.96, MonR2 \citep{rizzo2003} and S140 \citep{savage2004}. However, all the detections have been obtained after long integrations toward a single position and they lack the information on the spatial distribution.

The \MonR\ star-forming region, located at 830~pc (\citealt{herbst1976}), contains an ultracompact (UC)~\hii\ region surrounded by a series of PDRs with different physical conditions \citep{pilleri2013, trevino2014}. The main PDR, corresponding to IRS~1 (hereafter IF), is irradiated by a high UV field of $G_0>10^{5}$ (in units of the Habing flux; \citealt{habing1968}), and presents high densities ($>10^{5}$~cm$^{-3}$) and kinetic temperatures ($T_{\mathrm{k}}\approx 600$~K; \citealt{berne2009}). A second PDR, associated with the molecular peak MP2, is detected $40\arcsec$ north from IF, and shows chemical properties similar to those found in low- to mid-UV irradiated PDRs (\citealt{ginard2012}). Due to its proximity and physical conditions, \MonR\ turns to be an excellent candidate to study the \hi/H$_2$ interface. CO$^+$ is thought to be a good PDR tracer, and its distribution is potentially an excellent diagnostic tool to learn about the physical structure of these regions.

In this paper, we present a study of the CO$^+$ ($J=2$--1) transition line toward \MonR\ and compare its spatial distribution with \emph{Spitzer} data reported by \citet{berne2009}, \emph{Herschel} data from \citet{pilleri2012} and \citet{ossenkopf2013} and the HCO$^+$ and H$^{13}$CO$^+$ molecules from \citet{trevino2014}.  

\begin{figure*}[t]
\centering
\includegraphics[width=4.4cm]{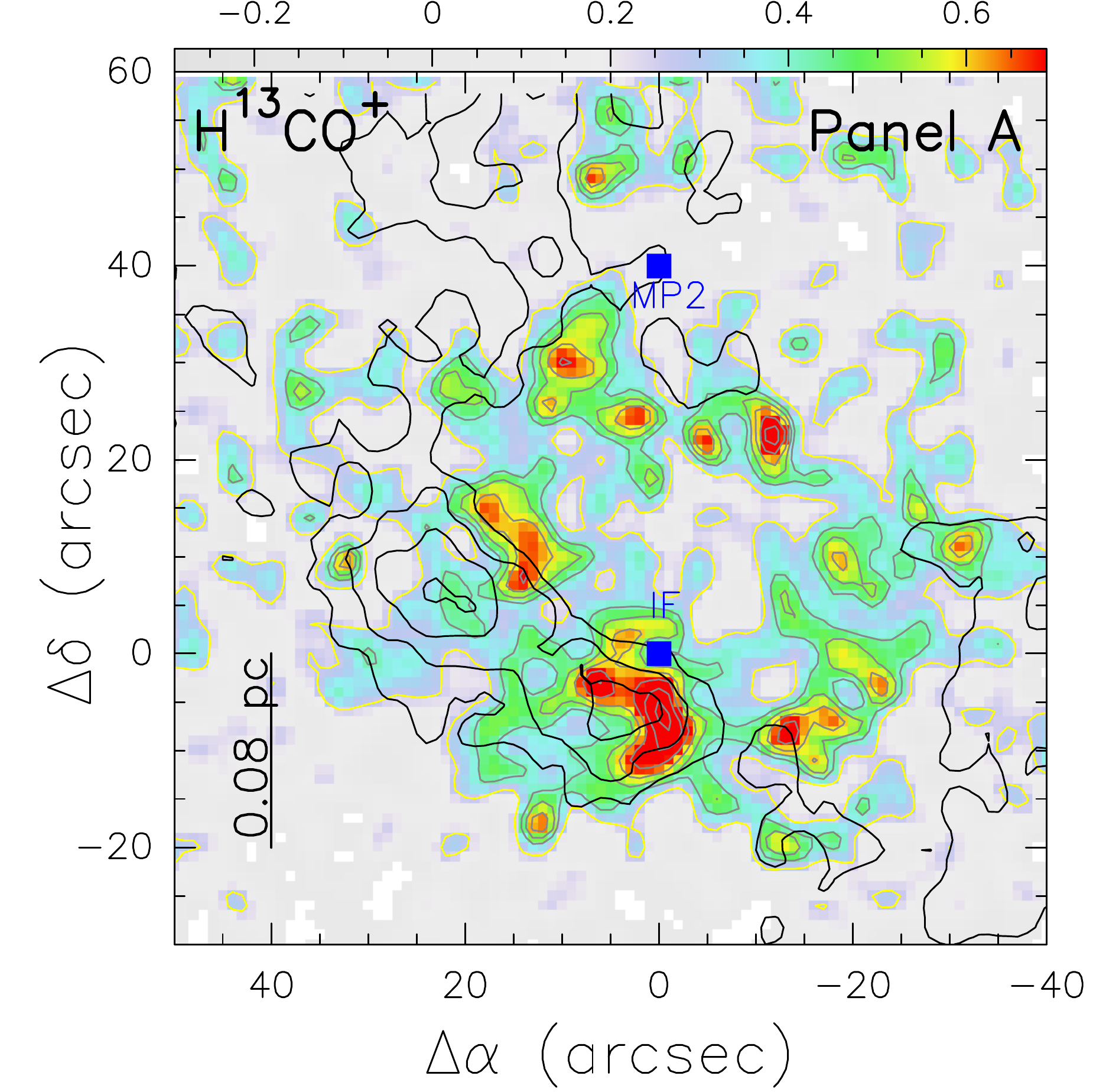}
\hspace{-0.3cm}\includegraphics[width=4.4cm]{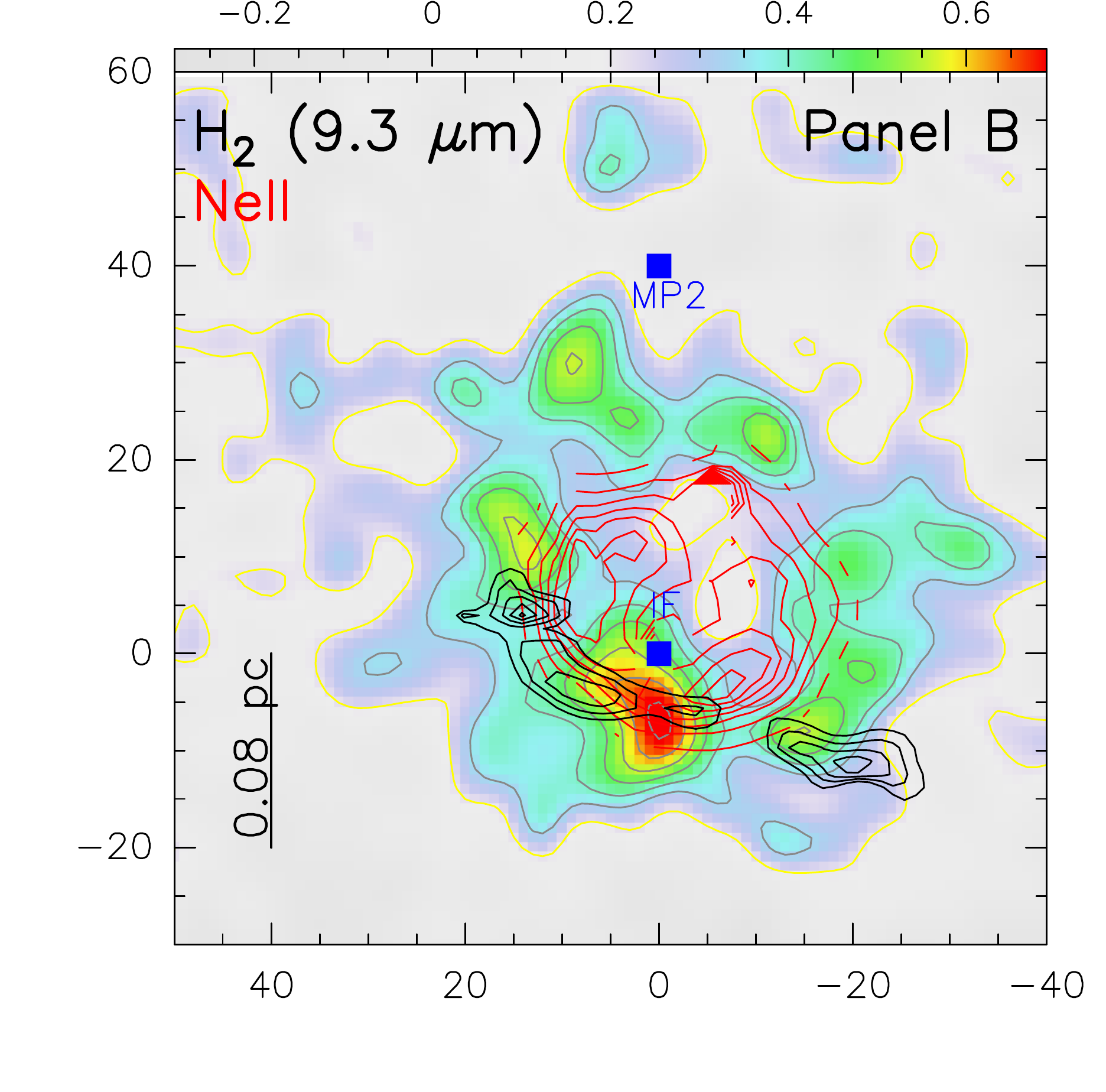}
\hspace{-0.3cm}\includegraphics[width=4.4cm]{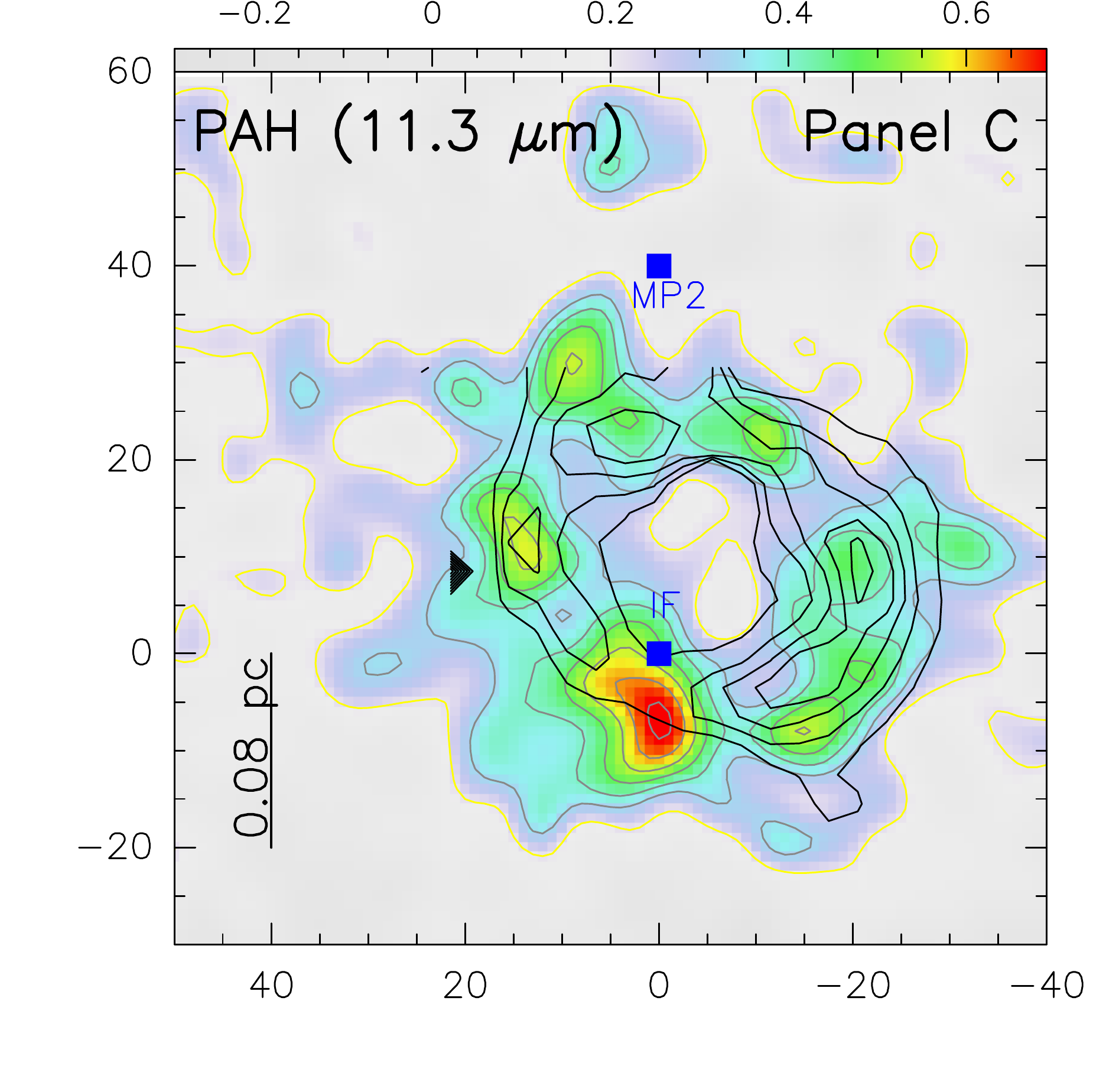}
\hspace{-0.3cm}\includegraphics[width=4.4cm]{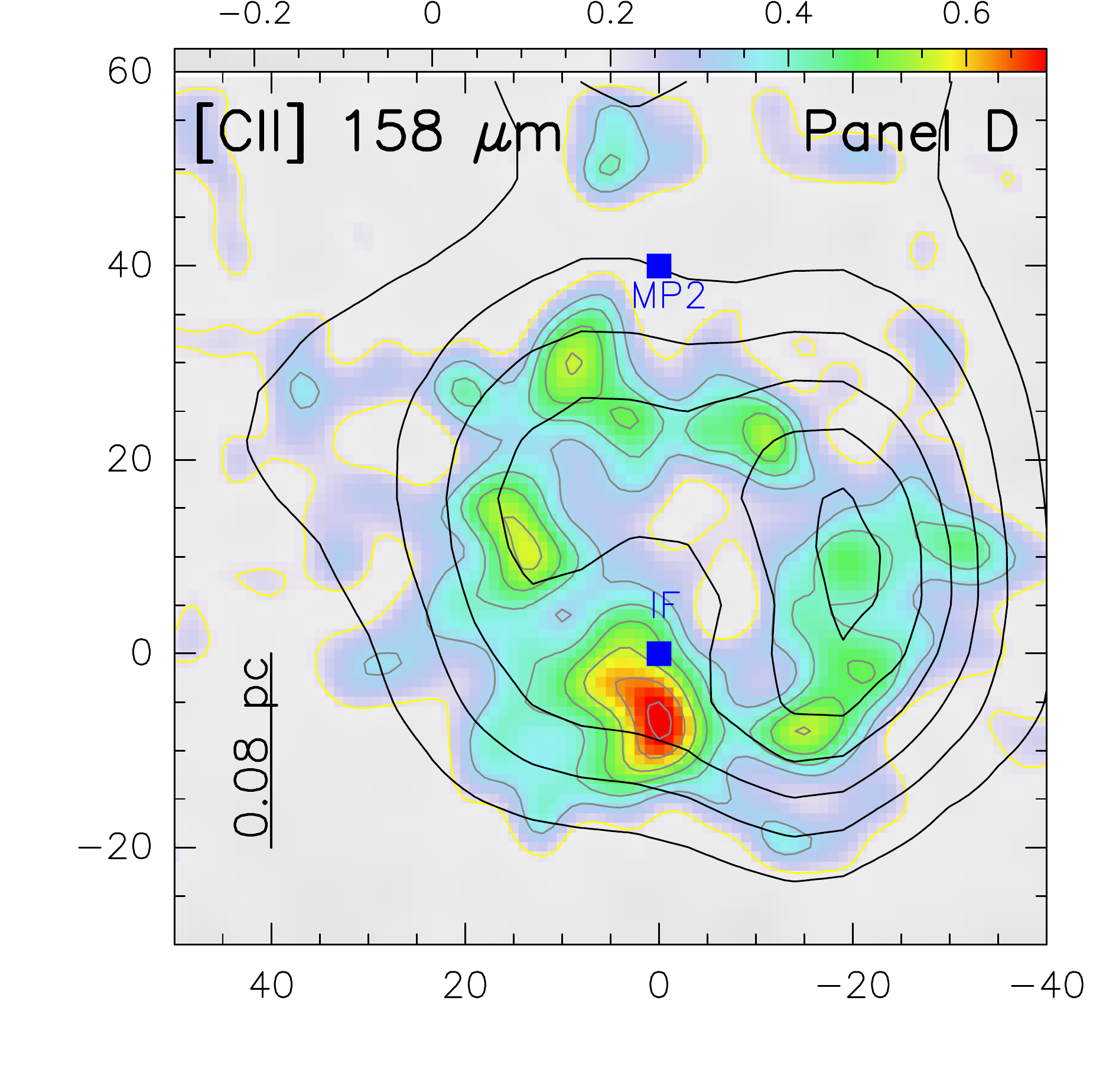}
\caption{Color image shows the integrated emission (in K~\kms) of the CO$^+$ line, with the original (11\arcsec, \emph{Panel A}) and smoothed (16\arcsec, \emph{Panel B} to \emph{D}) angular resolution. In \emph{Panels A} to \emph{D}, the gray contour levels range from 40\% to 100\%, in steps of 10\% of the intensity peak, where the lower contour level corresponds to a S/N$=5\sigma$. The yellow contour indicates the 3$\sigma$ emission. The blue squares indicate the IF and MP2 positions, where IF corresponds to $\alpha$(J2000)$=06\mathrm{h}07\mathrm{m}46.2\mathrm{s}$, $\delta$(J2000)$=-06^{\circ}23'08.3''$. \emph{Panel~A} shows the H$^{13}$CO$^+$\ (3--2) emission (black contours) tracing the molecular gas \citep{trevino2014}. \emph{Panel~B} shows the [Ne{\small II}] emission (red contours) tracing the \hii\ region and the emission of the H$_2$ S(3) rotational line at 9.7~$\mu$m (black contours). Black contours in \emph{Panel~C} show the PAHs (11.3~$\mu$m) emission. \emph{Panel~D} shows the [\ciii] emission at 158~$\mu m$ (black contours, \citealt{pilleri2014}). The \emph{Spitzer} data are explained in \citet{berne2009}.}
\label{figureA1}
\end{figure*}

\section{Observations and data reduction\label{s:obs}}

We observed $2'\times2'$ maps of CO$^+$ transition lines ($J=2$--1 at 235.380~GHz, 235.789~GHz and 236.062~GHz) using the IRAM-30m telescope (Pico Veleta, Spain). The observations were performed using the EMIR receiver with the Fast Fourier Transform Spectrometer (FTS) at 200~kHz of resolution. Throughout this paper, we use the main-beam brightness temperature ($T_{\mathrm{MB}}$) as intensity scale. The data were reduced using standard procedures with the CLASS/GILDAS package \citep{pety2005}. The three lines were detected, but only the 236.0625 GHz has sufficient signal-to-noise ratio (S/N) for good imaging. In order to improve the S/N, we smoothed the native observation to an angular resolution of 16\arcsec\ (see Fig.~\ref{figureA1}) and to a spectral resolution of 1~\kms. In the final datacube, CO$^+$ has linewidths of $6$--$8$~\kms\ and intensity peaks of 60--200~mK (with rms$\sim20$~mK). The main CO$^+$ line is located close to a bright $^{13}$CH$_3$OH line (at 236.0628~GHz). However, we dismiss the idea of a possible blending in the CO$^+$ line, as in the spectral line survey conducted toward \MonR\ \citep{trevino2016} we do not find $^{13}$CH$_3$OH emission at any frequency. Moreover, the main compound CH$_3$OH is not detected at the positions where CO$^+$ is bright. 

\section{Results\label{s:res}}

In Fig.~\ref{figureA1} we compare the CO$^+$ spatial distribution with other species. \emph{Panel~A} shows the H$^{13}$CO$^+$ (3--2) line emission (black contours) tracing the molecular gas \citep{trevino2014}, that is distributed around the CO$^{+}$ emission. CO$^+$ presents a clumpy structure, where the main CO$^+$ clumps seem to have a counterpart in the H$^{13}$CO$^{+}$ (3--2) emission. However, the peaks of CO$^+$ are located $\approx5$--$10$\arcsec\ closer to the \hii\ region, likely tracing an inner layer of the region. We find that CO$^{+}$ emission appears surrounding the \hii\ region with its intensity peak at the offset [0\arcsec, $-7$\arcsec], very close the IF position (see \emph{Panel~B} of Fig.~\ref{figureA1}). Moreover, the two most intense CO$^+$ clumps are correlated with the H$_2$ emission, in an area where the density is presumably larger. \emph{Panels~C} and \emph{D} show the PAHs and the [\ciii] emission, respectively (black contours). The CO$^+$ emission present a clumpy ring-like distribution, spatially coincident with the PAHs emission. The CO$^+$ secondary clumps are associated with the PAHs emission peaks, but not the most intense one. \emph{Panel~D} shows a comparison of the CO$^{+}$ spatial distribution with the emission of its chemical precursor [\ciii]. These species are spatially associated, with the main difference being the location of the peaks: CO$^+$ has its intensity peak to the south of the IF position, while the [\ciii] peak is located to the west. Therefore, we interpret that the CO$^+$ is found toward the densest area of the region (where all the molecular gas piles up, \eg\ the H$^{13}$CO$^+$ spatial distribution) as expected since the critical densities of H$^{13}$CO$^+$ ($n_{\mathrm{cr}}\sim 10^{5}$~cm$^{-3}$) and CO$^+$ ($n_{\mathrm{cr}}\sim$ a few $10^{5}$~cm$^{-3}$; \citealt{stauber-Bruderer2009}) are larger than that of [\ciii] ($n_{\mathrm{cr}}\sim$ a few $10^{3}$~cm$^{-3}$; \citealt{goldsmith2012}). It is worth noting that the CH$^+$ molecule is also related to the CO$^+$ and [\ciii] chemistry. When comparing their spatial distribution, we find that the CO$^{+}$ emission also coexist with CH$^+$ but, as [\ciii], its intensity peak is located to the west of the IF position \citep{pilleri2014}. 

\begin{figure*}[t]
\centering 
\includegraphics[width=6.4cm]{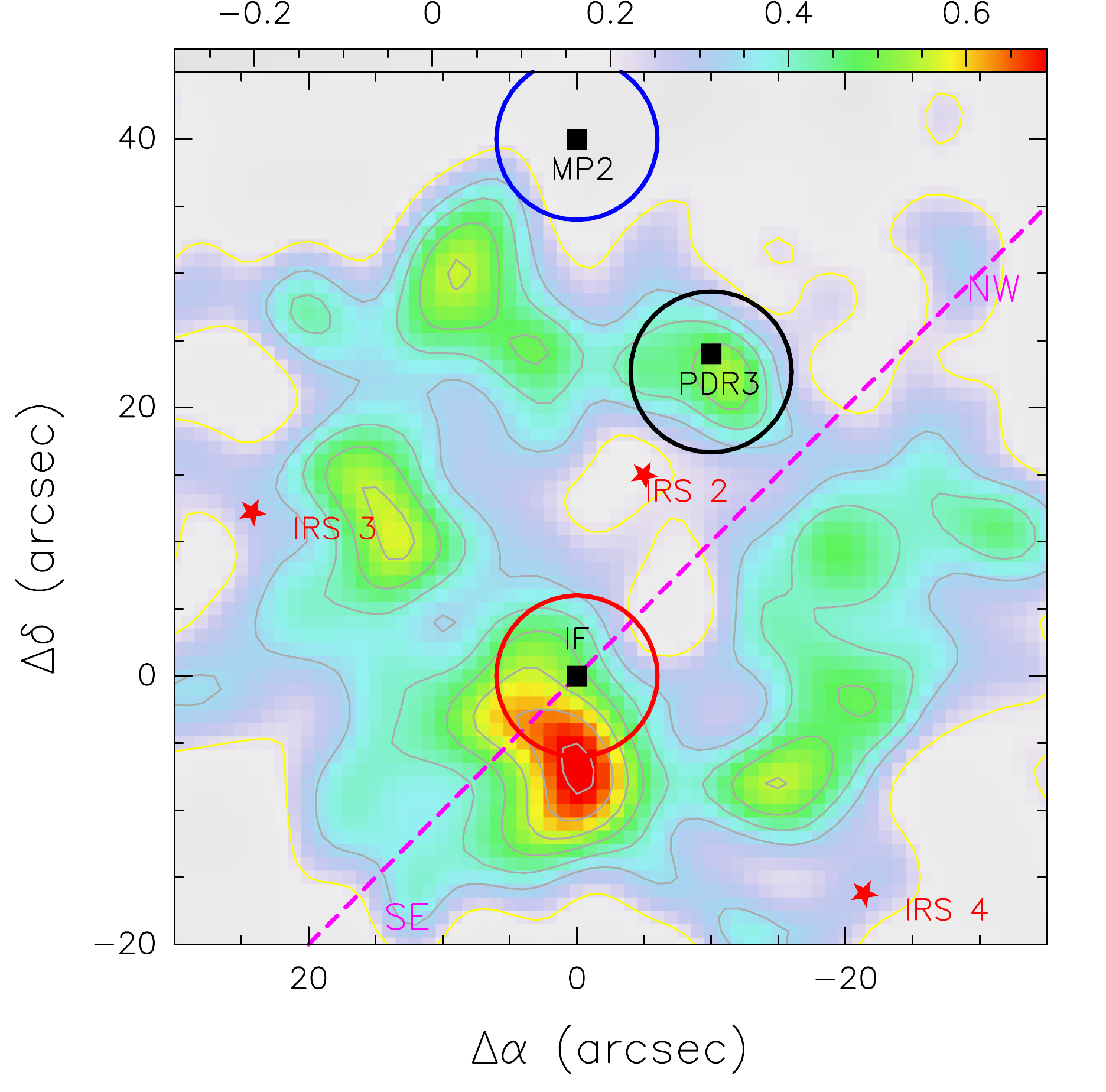} 
\hspace{0.5cm}\includegraphics[width=9.3cm]{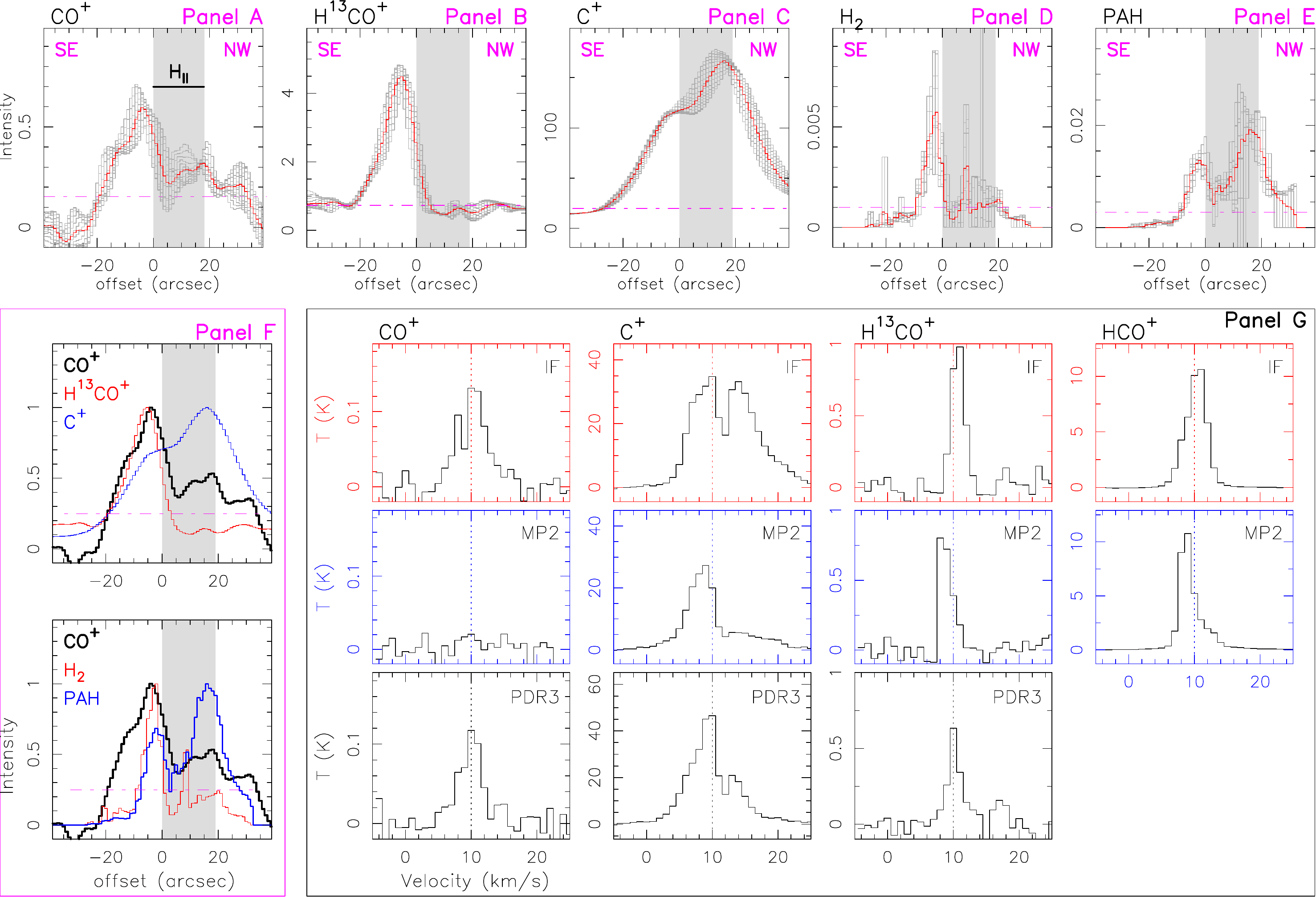}\\ 
\caption{\textbf{Left:} CO$^+$ integrated emission (see Fig.~\ref{figureA1}). The pink-dashed line  
indicates the direction of the intensity cuts. The color circles indicate the positions where spectra were extracted. \textbf{Right:} \emph{Panel~A} to \emph{E} show the intensity cuts (red continuum lines) of CO$^{+}$, H$^{13}$CO$^{+}$, [\ciii], H$_2$ and PAHs. In these panels, the gray continuum lines indicates the errors in the cuts. \emph{Panel~F} shows the comparison of the intensity cuts scaled to unity. In \emph{Panels A} to \emph{F}, the pink dot-dashed indicate the $3\sigma$ level for each species and the gray area indicates the position of the UC~\hii\ region. \emph{Panel~G} shows the CO$^+$, [\ciii], H$^{13}$CO$^+$ and HCO$^+$ spectra at IF, MP2 and PDR3. The color of the boxes are related to this positions (circles in left panel).} 
\label{figureA2}
\end{figure*}

In order to better understand the spatial distribution, we did intensity cuts with a position angle of 45\degr\ throughout the IF position and cutting the ring-like structure seen in the CO$^+$ emission at the south-east and north-west (pink dashed line in left panel of Fig.~\ref{figureA2}). The intensity cuts for the species CO$^+$, H$^{13}$CO$^+$, [\ciii], H$_2$ and PAHs are shown in \emph{Panels~A} to \emph{F} of Fig.~\ref{figureA2}. The molecular gas as traced by H$^{13}$CO$^+$ shows emission between the offset $-20$\arcsec and $0$\arcsec (corresponding to the area between [$20$\arcsec,$-20$\arcsec] and [$0$\arcsec,$0$\arcsec] in the map) with no emission associated with the \hii\ region. The most intense emission of CO$^+$ comes from this region, but there is also 3$\sigma$-level emission associated with the \hii\ region (gray-shaded area in the map). Note, however, that the CO$^+$ peak is closer to the \hii\ region than the H$^{13}$CO$^+$ peak. The [\ciii] emission is very intense and extended in the whole area, and its intensity cut shows two emission peaks: one coincident with the CO$^+$ peak, and a second one (the brightest one) on the opposite edge of the ring-like structure (to the north-west). The PAHs emission is weaker than [\ciii] but their intensity cuts are quite similar. Finally, the H$_2$ S(3) line only shows emission over $3\sigma$ between the offset $-10$\arcsec\ and 0\arcsec. Summarizing we can see a trend of spatial segregation, with the H$^{13}$CO$^+$ tracing the outer layers (far from the \hii\ region), then CO$^+$ and [\ciii] and PAHs peaking closer to the ionized gas, and finally the H$_2$ emission tracing a hotter layer close to the \hii\ region (\emph{Panel F} of Fig.~\ref{figureA2}).

The circles in the left panel of Fig.~\ref{figureA2} mark the positions where we compare the spectral profiles of CO$^{+}$, [\ciii], H$^{13}$CO$^{+}$ and HCO$^+$ (\emph{Panel~G}). The positions correspond to (i) the ionization front (IF at offset [0\arcsec, 0\arcsec], red), (ii) the offset [$-10$\arcsec, 24\arcsec] (close to the PDR3 in \citealt{berne2009}, black), and (iii) the MP2 position (offset [0\arcsec, 40\arcsec], blue) that is known to be a PDR with a low UV field and high density \citep{ginard2012}. In Fig.~\ref{figureA3}, we present a comparison of the CO$^+$ line profile (in red) with the [$^{13}$C{\small II}] \citep{ossenkopf2013}, $^{13}$CO (10--9) and CO (9--8) lines at IF \citep{pilleri2012}. We find that CO$^+$ presents larger linewidth than $^{13}$CO, H$^{13}$CO$^{+}$ and HCO$^+$ ($\approx7\pm0.7$~\kms\ vs 3--5~\kms). The CO$^+$ linewidth, however, is comparable to that of carbon recombination lines ($\approx$6~\kms; \citealt{trevino2016}). Similarly, the CO$^+$ line has a similar profile to that of the [$^{13}$C{\small II}] and the high excitation CO (9--8) line. The CO$^{+}$ and [$^{13}$C{\small II}] spectra are closely similar in the line shape and line widths; this similarity may be well explained if CO$^+$ lies in the dissociation front, between the ionized gas and the molecular gas. We find weak CO$^{+}$ emission toward the \hii\ region that might come from the back and front walls of the PDR.

On the basis of reactive ions (CH$^+$, OH$^+$, and H$_2$O$^+$) observations, \citet{pilleri2014} constructed a schematic view of the \MonR\ geometry. They find that the emission of the high density molecular gas seems to come from the back side (relative to the observer) of the \hii\ region. This is confirmed by the detection of the cluster at infrared wavelengths (\ie\ not obscured by in-front molecular gas). In this scenario, one expects that an expansion of the \hii\ region is indicated by an excess of red-shifted emission in the observed lines. However, while we find hints of the molecular tracers (H$^{13}$CO$^+$, CO) to be skewed toward red-shifted velocities, the PDR tracers (CO$^+$, [$^{13}$C{\small II}]) seem to have a spectral profile skewed toward blue-shifted velocities (Figs.~\ref{figureA2} and~\ref{figureA3}). In order to quantify this, we have measured the amount of emission in each spectral line that is blue-shifted and red-shifted with respect to the systemic velocity (10~\kms\ for IF and PDR3, and 8~\kms\ for MP2; \citealt{trevino2014}). For the IF position, we find $55\pm11$\% of the emission of PDR tracers to be blue-shifted, and only $33\pm6$\% for the molecular tracers. Similarly, for MP2 and PDR3 we find $44\pm9$\% and $55\pm11$\% of blue-shifted emission for the PDR tracers and $18\pm4$\% and $39\pm8$\% for the molecular tracers. In general, we have $51\pm10$\% of the PDR-tracer emission to be blue-shifted, and only $30\pm6$\% for the molecular tracers. Considering that the molecular gas is located behind the UC~\hii\ region, the difference between the PDR and molecular tracers can be explained if the PDR is formed by dense condensations that are being photo-evaporated. In this case, the photo-evaporated gas (PDR tracers) would be ejected toward us, and therefore, blue-shifted with respect to the molecular gas. This effect is also visible in Fig.~\ref{figureA3}, where [$^{13}$C{\small II}] presents its velocity peak at $\sim$9~\kms\ (blue-shifted) and the emission associated with the molecular gas is red-shifted. Finally, we do not find significant velocity gradients ($<1$~\kms) between the CO$^+$ clumps, suggesting low levels of turbulence between them, and consistent with the low expansion velocity found by \citet{fuente2010} and \cite{pilleri2012}.

\section{CO$^+$ and HCO$^+$ fractional abundances\label{s:abundances}}

On the basis of the spectra presented in Figs.~\ref{figureA2} and~\ref{figureA3}, we calculated the CO$^+$, HCO$^{+}$ and \cii\ column density ($N$) in ranges of 1~\kms\ at the three selected positions (see Appendix~\ref{a:abundances}). We assume optically thin emission and a beam filling factor of 1 for all the species. The $N$[CO$^+$] values have been calculated assuming a Boltzmann distribution of rotational levels with $T_{\mathrm{ex}}=18$~K. This value of the excitation temperature is based on calculations by \citet{stauber-Bruderer2009} and the detection of several CO$^+$ lines towards the Orion Bar (\citealt{cuadrado2015}, S.\ Cuadrado, priv.\ communication). We used the MADEX large velocity gradient (LVG; \citealt{cernicharo2012}) and RADEX code \citep{vandertak2007}, to derive $N$[H$^{13}$CO$^+$] and $N$[C$^+$]. According to the physical conditions derived by \citet{berne2009} from the H$_2$ ground state rotational lines, we assume $T_\mathrm{k}=600$~K, $n_\mathrm{H}=4\times 10^5$~cm$^{-3}$ for the IF and $T_\mathrm{k}=300$~K, $n_\mathrm{H}=4\times 10^4$~cm$^{-3}$ for PDR3. Unfortunately, the MP2 position is out of the \textit{Spitzer} map; for this position, we make a reasonable guess of $T_\mathrm{k}=300$~K and $n_\mathrm{H}=2\times 10^5$~cm$^{-3}$. The assumption of optically thin emission is not valid for the HCO$^+$\,(3--2) line at velocities close to cloud systemic velocity. Thus, between 8--11~\kms, we derived the $N$[HCO$^+$] using the rarer isotopologue line H$^{13}$CO$^+$\,(3--2) and assuming $^{12}$C/$^{13}$C$=50$ \citep{trevino2014}. The derived H$^{13}$CO$^+$ excitation temperatures ($\sim$10 $-$16 K) toward the IF and MP2 are in agreement with those measured by \citet{trevino2014}. The collisional rate coefficients for HCO$^+$ and H$^{13}$CO$^+$ are taken from \citet{flower1999}. To derived the $N$[C$^+$] at the IF position, between 6--12~\kms, we used the rarer isotopologue line [$^{13}$C{\small II}]. A significant fraction of the [\ciii] emission is expected to come from the atomic layer. The calculated C$^+$$-$H collisional rates are similar to those with H$_2$ within a factor 1.3 (\citealt{wiesenfeld2014}, \citealt{barinovs2005}). Hence, the relevant parameter regarding collisional excitation is the number of particles, either H or H$_2$. In our calculations we assume that the hydrogen is in molecular form which implies an uncertainty of a factor of 2 in the assumed density. Taking into account that we are well over the critical density of the [\ciii] 158~$\mu$m line, this translates into an uncertainty of $<$30\% in the $N$[\cii]. The obtained $N$[\cii] are in good agreement with those obtained by \citet{ossenkopf2013}.

We have computed the $N$[CO$^+$]/$N$[HCO$^+$] ratio. We find values between 0.01--0.1 toward IF, with the highest values in the velocity wings. Note that we have not detected CO$^+$ toward the PDR with lower UV field (MP2) with a significant upper limit of $N$[CO$^+$]/$N$[HCO$^+$]$<$0.008. Toward PDR3, we obtain values of $N$[CO$^+$]/$N$[HCO$^+$]$\sim 0.006$, \ie\ a factor of $\approx$2 lower than IF. C$^+$ is known to be a good probe of the skin ($A_\mathrm{v}<4$~mag) of PDRs. Because of the similar spatial distribution and velocity profiles between the CO$^+$ and \cii, we used C$^+$ to estimate the absolute fractional abundance of CO$^+$. We can safely assume that almost all the carbon is in C$^+$ in the region from which CO$^+$ is coming. Assuming a carbon elemental abundance of 10$^{-4}$ (\citealt{ossenkopf2013}, \citealt{wakelam2008}), we derive $X$[CO$^+$] between a few $10^{-11}$ to $\sim1.9\times 10^{-10}$ toward both, IF and PDR3. It is worth noting that the beam of CO$^+$, HCO$^{+}$, H$^{13}$CO$^{+}$, [$^{13}$C{\small II}] and [\ciii] are quite similar, thus the calculated $N$[CO$^+$]/$N$[HCO$^+$] and $X$[CO$^+$] values are not affected by beam filling factors. Our results are in good agreement with the model predictions presented by \citet{sternberg-Dalgarno1995}, for $G_0\sim 5\times 10^5$ and $n_{\mathrm{H}}\sim 10^6$~cm$^{-3}$. More recently, the models of \citet{spaans2007} predict the $X$[CO$^+$] in PDRs, for $n_\mathrm{H}=10^5$~cm$^{-3}$ and $G_0=10^{3.5}$. However, the high cosmic ray ionization rate (about 100 times larger than the Galactic value) prevents us from a direct comparison with \MonR. In general, the production of CO$^+$ seems to depend on the temperature of the gas. \citet{stauber-Bruderer2009} suggest that $X$[CO$^+$] of about 10$^{-11}$ are only reached in gas with $T_\mathrm{k}\ge300$~K. As $T_\mathrm{k}$ depends on $G_0$ and $n_{\mathrm{H}}$, it is expected that the production of CO$^+$ only occurs in regions with $n_{\mathrm{H}}\geq2\times 10^{4}$~cm$^{-3}$ and $G_0\ge10^{3}$. The densities and $G_0$ measured in IF and PDR~3 are in good agreement with these values, as IF presents $G_0\sim5\times 10^{5}$ and $n_{\mathrm{H}}\geq5\times 10^{4}$~cm$^{-3}$ (\citealt{rizzo2003}; \citealt{fuente2010}) and PDR3 presents $G_0\sim 3.7\times 10^{4}$ and $n_{\mathrm{H}}\sim 3.7\times 10^{4}$~cm$^{-3}$ (\citealt{berne2009}). 
 
\section{Discussion and summary\label{s:summary}}

We present a CO$^+$ map toward \MonR\ star-forming region. This is the first map ever reported of this reactive ion. The spatial distribution of CO$^+$ consists of a ring-like structure (similar to PAHs), tracing the layer between the \hii\ region and the molecular gas. The maps reveal a clumpy structure in the hot layer of the mainly atomic gas. Previous works (\citealt{young2000}; \citealt{goicoechea2016}) suggest that there exist fragmentation in the photodissociation front. This is, there are not uniform layers between the \hii\ region and the molecular cloud, but they present clumps that allow the radiation to penetrate deeper into the cloud. In this scenario where the PDR is conformed by a series of clumps, the emission of PDR tracers would be related to the external layers of dense clumps being photo-evaporated by the UV radiation. Despite the moderate angular resolution of our observations, we find hints that favor this scenario: (a) the spatial distribution of the CO$^+$ as observed in the higher-angular resolution (11$''$) map (\emph{Panel~A} of Fig.~\ref{figureA1}) suggests that the CO$^+$ emission is coming from the illuminated surface of the H$^{13}$CO$^{+}$ clumps, and b) the excess of blue-shifted emission seen for the PDR tracers in comparison with the molecular tracers. Considering that the molecular gas is located behind the UC~\hii\ region \citep{pilleri2014} and the chemical segregation, the difference in velocity between tracers can be explained if the PDR is formed by dense condensations that are being photo-evaporated. In this case, the photo-evaporated gas (PDR tracers) would be ejected toward us, and therefore, blue-shifted with respect to the molecular gas. Future higher angular resolution observations will help to confirm or discard this scenario. 

Finally, we have determined $X$[CO$^+$] in three positions. Toward IF we derive an abundance of a few $10^{-11}$, in agreement with chemical model predictions \citep{sternberg-Dalgarno1995} for $n_{\mathrm{H}}\sim 10^6$~cm$^{-3}$ and G$_{0}\sim5\times 10^5$. Toward MP2 we do not detect CO$^+$ emission with an upper limit to the CO$^+$ abundance of $<4\times 10^{-11}$. Abundances of $10^{-11}$--$10^{-10}$ had been previously observed in PDRs with $G_0>10^3$ Habing field (M17SW: \citealt{latter1993}, \citealt{stoerzer1995}; Orion Bar: \citealt{fuente1997}; NGC\,7023: \citealt{fuente2003}; G29.96$-$0.02: \citealt{rizzo2003}). The non-detection of CO$^+$ in MP2, together with the abundances found in the other PDRs, suggest that the production of CO$^+$ only occurs in dense regions with high radiation fields. High UV fields ($G_0>10^3$) and $n_{\mathrm{H}}$ ($>2\times 10^{4}$~cm$^{-3}$) are required to achieve gas temperatures $\geq300$~K that are necessary to produce high abundances of OH in the external layer of the PDR ($A_\mathrm{{V}}\sim1$~mag; \citealt{stauber-Bruderer2009}). This is also consistent with the non-detection of CO$^+$ in the Horsehead PDR \citep{goicoechea2009}, where the UV field is $G_0\sim100$ and presents chemical properties similar to MP2 \citep{ginard2012}. A counter-example that challenge this interpretation could be the detection of CO$^+$ toward S140 where the incident UV field is estimated to be $G_0\sim100$--300 \citep{savage2004}. However, the number of CO$^+$ detections is scarce and the statistics is not enough to draw firm conclusions between the relation of the CO$^+$ and the physical properties ($n_{\mathrm{H}}$ and $G_0$) of PDRs. A larger sample of objects need to be studied, including maps to characterize and understand the spatial distribution of the CO$^+$ in different environments.

\begin{acknowledgements}
SPTM, AF, JRG and JC thank the Spanish MINECO for funding support from grants AYA2012-32032, CSD2009-00038, FIS2012-32096, and ERC under ERC-2013-SyG, G. A. 610256 NANOCOSMOS. ASM and VO thank the Deutsche Forschungsgemeinschaft (DFG) for funding support via the collaborative research grant SFB 956, projects A6 and C1. PP acknowledges financial support from the Center National d’Etudes Spatiales (CNES).
\end{acknowledgements}

\begin{appendix}

\section{Comparison with other lines\label{a:comp}}

\begin{figure}[ht!]
\centering 
\includegraphics[width=8cm]{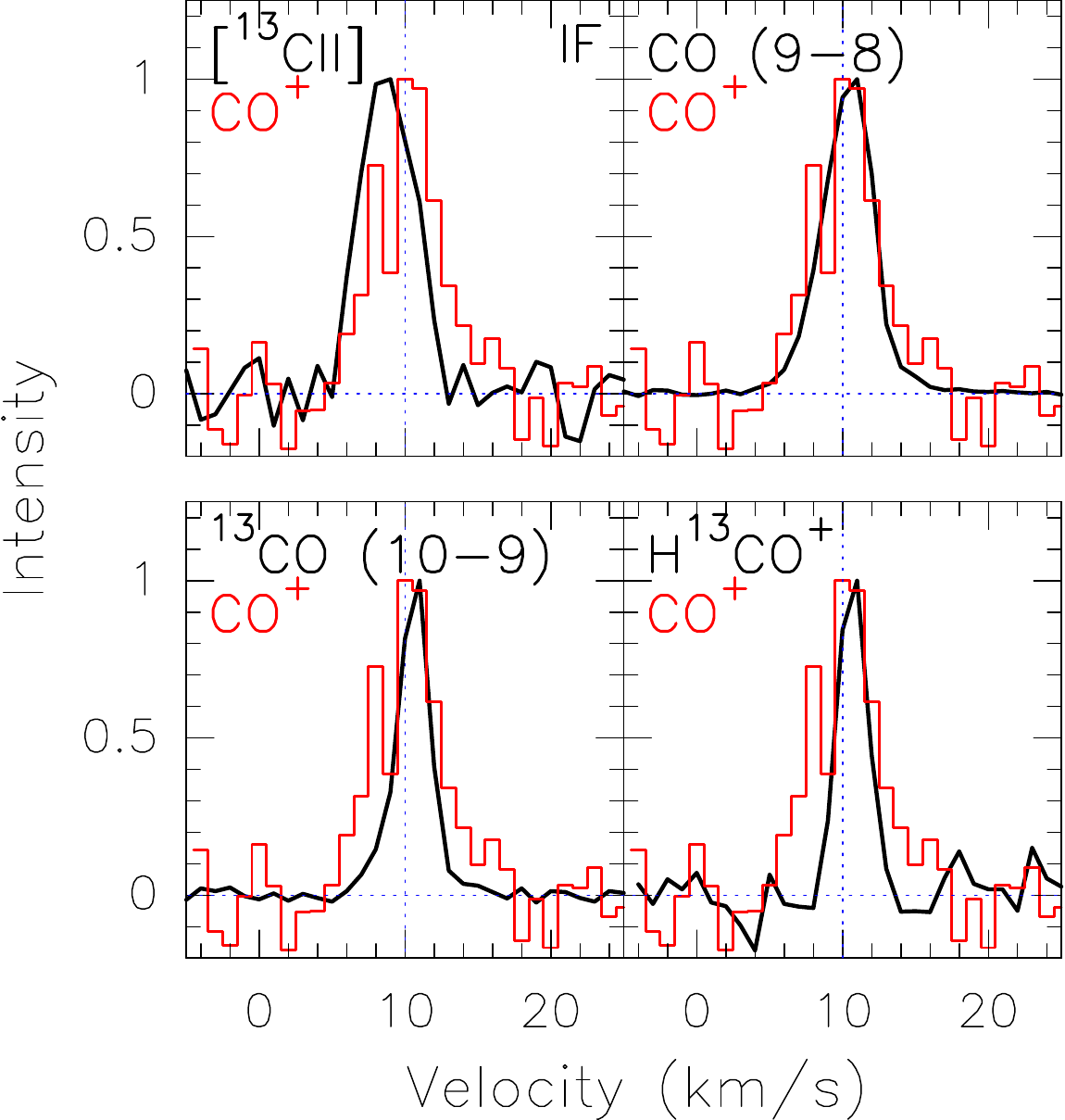}   
\caption{Comparison of the CO$^+$ line profile (in red) with the [$^{13}$C{\small II}], $^{13}$CO (10--9), CO (9--8) and H$^{13}$CO$^+$ lines (in black) at the IF position, with the intensity is scaled to unity. \emph{Herschel} data are presented in \citet{pilleri2012} and \citet{ossenkopf2013}. The blue dotted line indicates the systemic velocity (10~\kms).} 
\label{figureA3}
\end{figure}

In this section we present the comparison of the CO$^+$, [$^{13}$C{\small II}], $^{13}$CO (10--9) and CO (9--8) line profile. Figure~\ref{figureA3} shows the CO$^+$ (in red), [$^{13}$C{\small II}] \citep{ossenkopf2013}, $^{13}$CO (10--9) and CO (9--8) lines (in black) at the IF position \citep{pilleri2012}, with the intensity is scaled to unity. We find that CO$^+$ present larger linewidth than $^{13}$CO and H$^{13}$CO$^{+}$ ($\approx7\pm0.7$~\kms vs 3--5~\kms). Moreover, CO$^+$ line has a similar profile to that of the [$^{13}$C{\small II}] and CO line. The [$^{13}$C{\small II}] line presents its velocity peak at $\sim$9~\kms and the emission associated with the molecular gas (CO, $^{13}$CO) is red-shifted, while the systemic velocity is $\sim$10~\kms. 

\section{Column densities and fractional abundances\label{a:abundances}}

In this section we present the calculated CO$^+$, HCO$^{+}$ and \cii\ column densities ($N$), as well as the fractional abundance $X(\mathrm{CO}^+)$ of the IF, MP2 and PDR3 positions. The column densities were calculated in ranges of 1~\kms, from 5~\kms\ to 14~\kms. The CO$^+$ column densities have been calculated assuming a Boltzmann distribution of rotational levels with $T_{\mathrm{ex}}=18$~K. While for HCO$^+$ and \cii, we used RADEX and MADEX LVG codes (\citealt{vandertak2007} and \citealt{cernicharo2012}), considering the physical conditions derived by \citet{berne2009}. It is worth noting that for the IF position the HCO$^{+}$ emission is optically thick, thus we used the rarer isotopologue H$^{13}$CO$^{+}$ to correct the opacity effects and derive $N$[HCO$^+$] in each position, assuming $^{12}$C/$^{13}$C$=50$ \citep{trevino2014}. Regarding the $N$[\cii] estimation, we note that [\ciii] is optically thick towards the IF position (as seen by comparing the [\ciii] and [$^{13}$C{\small II}] lines): under the assumption of optically thin emission, [$^{13}$C{\small II}] results in a column density a factor 3--5 larger than that derived from the main isotopologue [\ciii]. From this, and considering that the other two positions are associated with less dense gas, we can infer that the assumption of optically thin emission applied to the [\ciii] in MP2 and PDR3, may underestimate the column density by a factor $<3$. Because of the similar spatial distribution and velocity profiles between the CO$^{+}$ and \cii\, we used $N$[\cii] to estimate the absolute fractional abundance $X(\mathrm{CO}^+)$. For the calculations, we consider a beam filling factor of 1. This assumption is consistent with the fact that the beams of [\ciii] at 158$\mu$m, H$^{13}$CO$^{+}$ (3--2), HCO$^{+}$ (3--2) and CO$^{+}$ (2--1) are quite similar. Thus the calculated $N$[CO$^{+}$]/$N$[HCO$^{+}$] and $X$[CO$^{+}$] values are not affected by the beam filling factor. Table~\ref{Table1} lists the calculated values of $N$[CO$^+$], $X$[CO$^+$], $N$[HCO$^+$], $N$[\cii] and $N$[CO$^+$]/$N$[HCO$^+$] in every velocity range.

\begin{table*}
\centering
\footnotesize
\caption{Column densities and ratios, in ranges of 1~\kms, of the selected positions (see Fig.~\ref{figureA2}). We assumed LTE and $T_{\mathrm{ex}}=18$~K to calculate $N$[CO$^+$], while $N$[HCO$^+$] and $N$[\cii] have been calculated assuming LVG. For the IF position, we assume $T_\mathrm{k}=600$~K, $n_\mathrm{H}=4\times 10^5$~cm$^{-3}$. For MP2 position, we assume $T_\mathrm{k}=300$~K and $n_\mathrm{H}=2\times 10^5$~cm$^{-3}$. For PDR3 position, we assume $T_\mathrm{k}=300$~K and $n_\mathrm{H}=4\times 10^4$~cm$^{-3}$.} 
\begin{tabular}{l l | c| c c c c c c c | c |r} 
\hline 
\noalign{\smallskip} 
Velocity range 
&(\kms)
&5 -- 6 
&6 -- 7 
&7 -- 8 
&8 -- 9 
&9 -- 10
&10 -- 11
&11 -- 12
&12 -- 13
&13 -- 14
&Total$^{*}$    
\\ 
\hline
For the CO$^+$ line
&
&$\sim3\sigma$ 
& 
& 
& 
&$>3\sigma$
&
&
&
&$\sim3\sigma$
& 
\\
\hline
\noalign{\smallskip}
\multicolumn{2}{l}{IF position --- offset [$0\arcsec$, $0\arcsec$]}
&\multicolumn{7}{l}{} \\
\hline
\noalign{\smallskip}
$N$[CO$^+$]              &(in $10^{11}$~cm$^{-2}$)       &\phs0.14      &\phs0.50       &\phs1.34       &\phs1.34      &\phs1.20        &\phs1.90        &\phs1.60            &\phs0.82            &\phs0.57            &\phn9.75$^{a}$         \\
$N$[HCO$^+$]             &(in $10^{13}$~cm$^{-2}$)       &$<0.04$       &\phs0.06       &\phs0.15       &\phs0.34      &\phs1.44        &\phs2.42        &\phs1.92            &\phs0.70            &\phs0.05            &\phn6.79$^{a}$        \\
$N$[C$^+$]               &(in $10^{17}$~cm$^{-2}$)       &\phs0.63      &\phs4.00       &\phs5.10       &\phs6.00      &\phs5.60        &\phs4.70        &\phs3.30            &\phs1.24            &\phs1.67            &48.00$^{b}$        \\
\\
$N$[CO$^+$]/$N$[HCO$^+$] &                               &$>0.04$       &\phs0.08       &\phs0.09      &\phs0.04      &\phs0.08        &\phs0.08        &\phs0.01            &\phs0.01            &\phs0.11            &\phn0.02        \\  
$X$[CO$^+$]              &(in $10^{-11}$)                &\phs2.22      &\phs1.47       &\phs2.16      &\phs2.23      &\phs2.14        &\phs4.04        &\phs4.90            &\phs6.61            &\phs3.41            &\phn2.03        \\
\hline
\noalign{\smallskip}
\multicolumn{2}{l}{MP2 position --- offset [$0\arcsec$, $40\arcsec$]}
&\multicolumn{7}{l}{} \\
\hline
\noalign{\smallskip}
$N$[CO$^+$]              &(in $10^{11}$~cm$^{-2}$)       &$<1.71$      &$<1.71$     &$<1.71$      &$<1.71$     &$<1.71$    &$<1.71$     &$<1.71$    &$<1.71$     &$<1.71$       &$<4.5$$^{d}$   \\
$N$[HCO$^+$]             &(in $10^{13}$~cm$^{-2}$)       &$<0.04$     &\phs0.06       &\phs0.29        &\phs2.21       &\phs1.46      &\phs0.71       &\phs0.29      &\phs0.06       &\phs0.06         &\phs\phn5.84$^{a}$ \\
$N$[C$^+$]               &(in $10^{17}$~cm$^{-2}$)       &\phs0.63    &\phs0.97       &\phs1.31        &\phs1.52       &\phs1.38      &\phs0.78       &\phs0.34      &\phs0.29       &\phs0.33         &\phs10.15$^{c}$    \\ 
\\
$N$[CO$^+$]/$N$[HCO$^+$] &                               &$>0.043$    &$<0.285$   &$<0.058$    &$<0.007$   &$<0.011$  &$<0.024$   &$<0.058$  &$<0.29$   &$<0.285$             &$<0.008$ \\  
$X$[CO$^+$]              &(in $10^{-10}$)                &$<2.71$    &$<1.76$    &$<1.31$     &$<1.13$    &$<1.24$   &$<2.19$    &$<5.03$   &$<5.89$    &$<5.18$      &$<0.45$                 \\    
\hline
\noalign{\smallskip}
\multicolumn{2}{l}{PDR3 position --- offset [$-10\arcsec$, $24\arcsec$]}
&\multicolumn{7}{l}{} \\
\hline
\noalign{\smallskip}
$N$[CO$^+$]              &(in $10^{11}$~cm$^{-2}$)       &$<1.35$    &\phs0.58  &\phs1.19       &\phs1.77   &\phs2.45   &\phs2.68     &\phs1.66    &\phs0.72        &$<1.35$          &13.02$^{a}$  \\
$N$[HCO$^+$]             &(in $10^{13}$~cm$^{-2}$)       &$<3.34$    &$<3.34$   &$<3.34$        &\phs1.70   &\phs5.17   &\phs6.10     &\phs3.04    &$<3.34$         &$<3.34$          &20.22$^{a}$ \\
$N$[C$^+$]               &(in $10^{17}$~cm$^{-2}$)       &\phs0.99   &\phs1.37  &\phs1.68       &\phs2.26   &\phs2.70   &\phs1.93     &\phs0.95    &\phs1.00        &\phs1.14         &18.04$^{c}$  \\ 
\\
$N$[CO$^+$]/$N$[HCO$^+$] &                               &$>0.004$   &$>0.002$  &$>0.004$       &\phs0.010  &\phs0.005  &\phs0.004    &\phs0.005   &$>0.002$        &$>0.004$         &\phn0.006 \\  
$X$[CO$^+$]              &(in $10^{-10}$)                &\phs1.36   &\phs0.42  &\phs0.71       &\phs0.78   &\phs0.91   &\phs1.92     &\phs1.75    &\phs0.72        &\phs1.18         & 0.72\phn\\    
\hline
\end{tabular} 
\begin{list}{}{}{{}}
\item $^{*}$ Total velocity range of each line. $^{a}$ In a velocity range of 5 -- 14~\kms. $^{b}$ In a velocity range of 4 -- 25~\kms. $^{c}$ In a velocity range of 0 -- 30~\kms.
$^{d}$ considering the \emph{rms} in a velocity range of 7~\kms.  
\end{list}
\label{Table1}
\end{table*}

\end{appendix}

\begin{thebibliography}{}

\bibitem[Bern{\'e} et al.(2009)]{berne2009} Bern{\'e}, O., Fuente, A., Goicoechea, J.~R., et al.\ 2009, \apjl, 706, L160  
\bibitem[Barinovs et al.(2005)]{barinovs2005} Barinovs, {\u G}., \et\ 2005, \apj, 620, 537
\bibitem[Cernicharo (2012)]{cernicharo2012} Cernicharo, J. 2012, in EAS Publ. Ser., 58, 251
\bibitem[Cuadrado et al.(2015)]{cuadrado2015} Cuadrado, S., Goicoechea, J.~R., Pilleri, P., et al.\ 2015, \aap, 575, A82 
\bibitem[Flower (1999)]{flower1999} Flower D.~R. 1999, MNRAS, 305, 651
\bibitem[Fuente \& Mart{\'{\i}}n-Pintado(1997)]{fuente1997} Fuente, A., \& Mart{\'{\i}}n-Pintado, J.\ 1997, \apjl, 477, L107 
\bibitem[Fuente et al.(2003)]{fuente2003} Fuente, A., Rodr{\i}guez-Franco, A., et al.\ 2003, \aap, 406, 899  
\bibitem[Fuente et al.(2010)]{fuente2010} Fuente, A., Bern{\'e}, O., Cernicharo, J., et al.\ 2010, \aap, 521, L23 
\bibitem[Ginard et al.(2012)]{ginard2012} Ginard, D., Gonz{\'a}lez-Garc{\'{\i}}a, M., Fuente, A., et al.\ 2012, \aap, 543, A27 
\bibitem[Goicoechea et al.(2009)]{goicoechea2009} Goicoechea, J.~R., Pety, J., Gerin, M., et al.\ 2009, \aap, 498, 771 
\bibitem[Goicoechea et al.(2016)]{goicoechea2016} Goicoechea, J.R. , Pety. J. Cuadrado et al. 2016 submitted. 
\bibitem[Goldsmith et al.(2012)]{goldsmith2012} Goldsmith, P.~F., Langer, W.~D., et al.\ 2012, \apjs, 203, 13 
\bibitem[Habing(1968)]{habing1968} Habing, H.~J.\ 1968, \bain, 20, 120 
\bibitem[Herbst \& Racine(1976)]{herbst1976} Herbst, W., \& Racine, R.\ 1976, \aj, 81, 840  
\bibitem[Latter et al.(1993)]{latter1993} Latter, W.~B., Walker, C.~K., \& Maloney, P.~R.\ 1993, \apjl, 419, L97 
\bibitem[Ossenkopf et al.(2013)]{ossenkopf2013} Ossenkopf, V., R{\"o}llig, M., Neufeld, D.~A., et al.\ 2013, \aap, 550, A57 
\bibitem[Pety et al.(2005)]{pety2005} Pety, J., Teyssier, D., Foss{\'e}, D., et al.\ 2005, \aap, 435, 885  
\bibitem[Pilleri et al.(2012)]{pilleri2012} Pilleri, P., Fuente, A., Cernicharo, J., et al.\ 2012, \aap, 544, A110 
\bibitem[Pilleri et al.(2013)]{pilleri2013} Pilleri, P., Trevi{\~n}o-Morales, S., Fuente, A., et al.\ 2013, \aap, 554, A87 
\bibitem[Pilleri et al.(2014)]{pilleri2014} Pilleri, P., Fuente, A., Gerin, M., et al.\ 2014, \aap, 561, A69  
\bibitem[Rizzo et al.(2003)]{rizzo2003} Rizzo, J.~R., Fuente, A., et al.\ 2003, \apjl, 597, L153  
\bibitem[Spaans \& Meijerink(2007)]{spaans2007} Spaans, M., \& Meijerink, R.\ 2007, \apjl, 664, L23 
\bibitem[Savage \& Ziurys(2004)]{savage2004} Savage, C., \& Ziurys, L.~M.\ 2004, \apj, 616, 966  
\bibitem[St{\"a}uber \& Bruderer(2009)]{stauber-Bruderer2009} St{\"a}uber, P., \& Bruderer, S.\ 2009, \aap, 505, 195
\bibitem[Sternberg \&\ Dalgarno(1995)]{sternberg-Dalgarno1995} Sternberg, A., \&\ Dalgarno, A.\ 1995, \apjs, 99, 565 
\bibitem[Stoerzer et al.(1995)]{stoerzer1995} Stoerzer, H., Stutzki, J., \& Sternberg, A.\ 1995, \aap, 296, L9  
\bibitem[Trevi{\~n}o-Morales et al.(2014)]{trevino2014} Trevi{\~n}o-Morales, S.~P., Pilleri, P., Fuente, A., et al.\ 2014, \aap, 569, A19 
\bibitem[Trevi{\~n}o-Morales (2016)]{trevino2016} Trevi{\~n}o-Morales, S.~P., 2016, PhD Thesis  
\bibitem[van der Tak et al.(2007)]{vandertak2007} van der Tak, F.~F.~S., Black, J.~H., Sch{\"o}ier, F.~L. et al.\ 2007, \aap, 468, 627 
\bibitem[Wakelam \& Herbst(2008)]{wakelam2008} Wakelam, V., \& Herbst, E.\ 2008, \apj, 680, 371-383 
\bibitem[Wiesenfeld \& Goldsmith(2014)]{wiesenfeld2014} Wiesenfeld, L., \& Goldsmith, P.~F.\ 2014, \apj, 780, 183 
\bibitem[Young et al.(2000)]{young2000} Young Owl, R.~C., et al.\ 2000, \apj, 540, 886
\end{thebibliography}
\end{document}